# Triplet State and Auger-Type Excitation Originating from Two-Electron Tunneling in Field Emission Resonance on Ag(100)

Shin-Ming Lu, Ho-Hsiang Chang, Wei-Bin Su,* Wen-Yuan Chan, Kung-Hsuan Lin, and Chia-Seng Chang

*Institute of Physics, Academia Sinica, Nankang, Taipei 11529, Taiwan*

In this study, we discovered that the energy gap above the vacuum level in the projected bulk band structure of Ag(100) prevents electrons in the first-order field emission resonance (FER) from inducing the surface plasmons. This mechanism allows light emission from FER to reveal characteristics of triplet states and Auger-type excitation resulting from two-electron tunneling in FER. According to optical spectra, surface plasmons can be induced by electrons in the zeroth-order FER. However, corresponding radiative decay can also trigger Auger-type excitation, whose energy state is influenced by the sharpness-dependent image potential acting on the scanning tunneling microscope tip.

Quantum tunneling plays a crucial role in phenomena such as radioactive α decay [1], nuclear fusion in stars [2], field emission [3], and Josephson effect [4]. It also has major applications in modern technologies, such as tunnel diodes [5], scanning tunneling microscope (STM) [6], and quantum computers [7]. According to the general quantum tunneling theory, no correlation occurs between tunneling particles (i.e., one-particle tunneling). For example, field emission can be described by the Fowler-Nordheim theory [8], which is based on one-electron tunneling. Therefore, field emission resonance (FER) [9-30] was thought to follow one-electron tunneling. However, a recent study has indicated that the linewidth of FER observed on $MoS_2$ surfaces may vary by up to one order of magnitude [31] as a result of two-electron tunneling into the quantized state of FER through the exchange interaction [32,33]. Two-electron tunneling is a rare phenomenon that is observed only in unique physical systems, such as superconducting tunnel junction [4,34–36]. Exchange interaction between electrons typically occurs in ferromagnetic and antiferromagnetic materials. This two-electron tunneling in FER formed under the STM configuration is worth exploring.

The quantized state in FER is occupied by two electrons of opposite spin through two-electron tunneling. Because FER electrons are excited electrons, light emission from FER had been observed [37,38]. In this study, we investigated whether light emission of FER provides information regarding the interplay of paired electrons. According to a recent study on Ag(100), the formation probability of electron pairs can be considerably increased [39] because Ag(100) has an energy gap above the vacuum level [40]. Therefore, Ag(100) should be an ideal system for examining such interplay. Overall, our optical results revealed two types of interplay: Auger-type excitation and spin-spin interaction between paired electrons. In Auger-type excitation, one electron is excited to a higher energy state by photons resulting from the radiative decay of surface plasmons [41–51] induced by another electron or by a photon directly emitted from another electron, depending on whether STM bias voltage is set to zeroth or first-order FER energy. In spin-spin interaction between the paired electrons, spin flipping occurs, and a triplet state is formed. The light emitted by these two types of interplay reflects the transition of electrons between quantized states, which has never been investigated in previous studies on FER light emission.

In this experiment, clean Ag(100) surfaces were prepared using ion sputtering followed by annealing at 600 °C over several cycles. The sample was then transferred to ultra-high-vacuum STM operated at 5 K. PtIr tips were used to perform Z-V spectroscopy at a current of 50 nA. FER was then examined using either a lock-in technique, in which a dither voltage of 30 mV was added at a frequency of 1 kHz to the bias voltage, or a numerical differentiation for the Z-V spectrum. The collected light was guided to a multimode fiber in the air, which was connected to an optical spectrometer with liquid nitrogen cooling.

To observe the relaxation of electrons in the quantized state of FER (hereinafter referred to as the FER state), the bias voltage was set to the FER energy. Because the FER energy varies as the sharpness of the STM tip changes [38,52,53], the FER spectra were examined before and after the acquisition of the optical spectrum to ensure that FER energies did not change during the light collection process [Fig. 1(a)]. The numbers in Fig. 1(a) denote the orders of the FER peaks. The bias voltages were set to the energies of zeroth-order FER (FER0) and first-order FER (FER1) because both fall within the energy gap [40]. Figure 1(b) displays the optical spectrum of the bias voltage set to the energy of FER1 ($E_1$). In addition to a peak with a considerably low intensity at 3 eV (peak c in the inset), a pair of peaks (peaks a and b) can be observed between photon energies of 1.3 to 2.0 eV. However, when the bias voltage was set to the energy of FER0 ($E_0$), the optical spectrum depicted in Fig. 1(c) exhibited a substantial change, wherein the intensity distribution below 2.7 eV was replaced by a broad distribution (highlighted in the inset), and peak c became substantially more pronounced, indicating the different relaxation behaviors for electrons between FER1 and FER0 states.

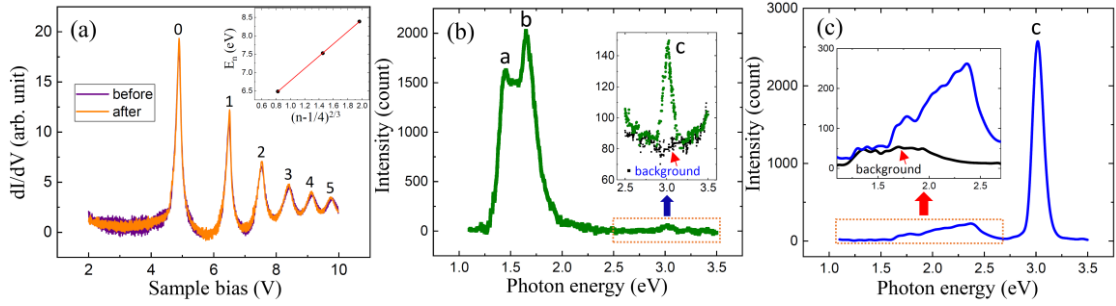

FIG. 1 (a) FER spectra before and after light emission measurement. Inset: plot of FER energy $E_n$ versus $(n-1/4)^{2/3}$. (b) Optical spectrum of bias voltage set to the energy of FER1 in (a). Inset: spectrum and spectrometer background at an energy range of 2.5–3.5 eV. (c) Optical spectrum of bias voltage set to the energy of FER0 in (a). Inset: spectrum and spectrometer background at an energy range of 1.1–2.7 eV.

According to a previous STM study, luminescence from metallic quantum wells can reveal peak features in the optical spectra because of the transition of electrons between quantum well states [54]. Similarly, the peak signals depicted in Fig. 1(b) were ascribed to the transition of electrons between FER states, which does not explain the twin-peak feature having a splitting energy of 0.19 eV. We suggest that this splitting is due to spin-spin interaction. In field emission, electrons are emitted separately because of one-electron tunneling. However, in FER, when an electron is field-emitted into FER1 state [Fig. 2(a)], because it reflects off the surface, its wave function overlaps with that of another electron with an opposite spin in the STM tip, thereby inducing exchange interaction. Consequently, these two electrons should be described by an antisymmetrical total wave function. Because their spins are opposite, their total space wave function is symmetrical. For a symmetrical wave function, electrons at the same location have the highest probability density, indicating that an up-spin electron can attract the emission of a down-spin electron through the exchange interaction, and vice versa. Therefore, FER involves two-electron tunneling, which causes the FER1 state to be occupied by two electrons of opposite spin per unit time [Fig. 2(b)].

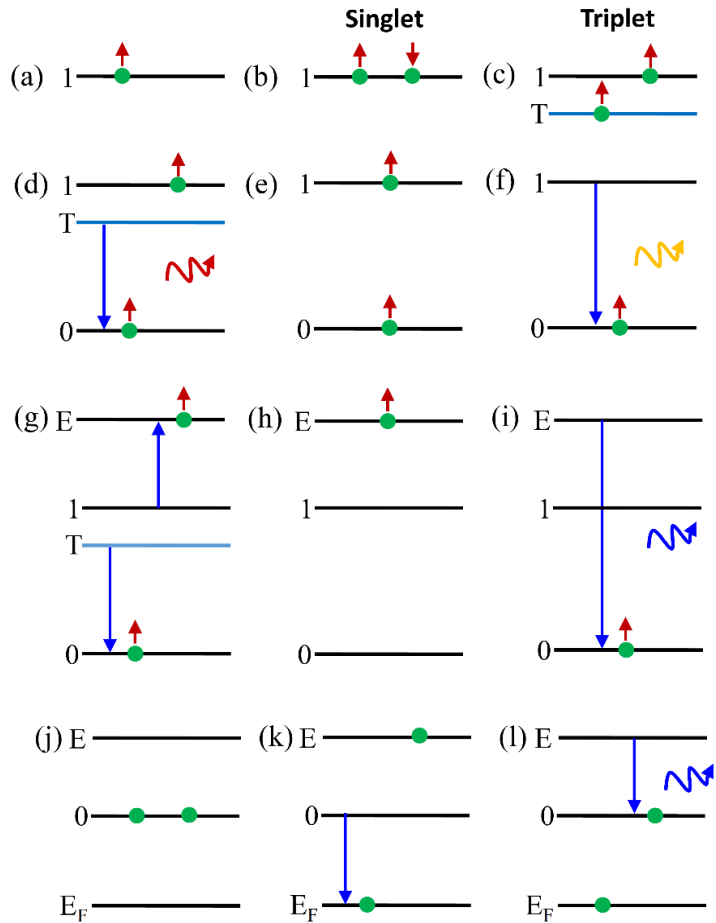

FIG. 2 Step-by-step illustrations (not to scale) of light emission from a triplet state [(a)-(f)] and Auger-type excitation [(g)-(l)] (see text).

These two electrons in the vacuum may have spin-spin interaction to form the singlet state. However, their energy may be lower if they form a triplet state. Therefore, the most likely scenario is that below FER1 state, an energy state called the T state exists. As shown in Fig. 2(c), one of two electrons jumps to the T state and its spin flips to form a triplet state. The electron in the T state then jumps to the FER0 state by emitting a photon [Fig. 2(d)], and T state vanishes [Fig. 2(e)] because the FER1 state becomes occupied by only one electron. Because both electrons have the same spin, the electron in the FER1 state emits another photon only after the electron that is temporarily trapped in the FER0 state due to quantum trapping [31,40] exits [Fig. 2(f)]. In the presence of a triplet state, these two photons may have different energies, resulting in a twin-peak feature. For peak c in Fig. 1(b), we suggest that the transition energy of the electron from the T state to the FER0 state is transferred to excite the FER1 electron to an excited state called the E state [Fig. 2(g)]. After the FER0 electron exits [Fig. 2(h)], the electron transition process from the E state to the FER0 state releases a photon [Fig. 2(i)]. The emergence of peak c indicates that in addition to the individual emission of two photons, there is also an alternative pathway similar to Auger-type excitation in the relaxation process of paired electrons. However, the probability of this pathway is considerably low, resulting in substantially low intensity in peak c.

In Fig. 1(c), peak c indicates that Auger-type excitation may occur when the bias voltage is set at $E_0$. However, below the FER0 state, no quantized state is available for electrons to jump and release photons. How, then, are electrons in the FER0 state excited to the E state? We suggest that the source is radiative decay of surface plasmons [41–51], which is represented by a broad distribution between 1.5 and 2.5 eV in Fig. 1(c). According to Leon et al. [55], the radiative decay of surface plasmon induced by a single electron can result in two photons. As shown in Fig. 1(c), because photon energy can reach up to 2.5 eV, the total energy of two photons is sufficient to excite an electron to the E state. The following is the detailed process. One of the paired electrons occupying the FER0 state [Fig. 2(j)] generates surface plasmons through inelastic scattering. As a result of surface plasmon decay, two photons are emitted and excite another electron to the E state [Fig. 2(k)]. The electron transition process from the E state to the FER0 state then releases a photon [Fig. 2(l)], represented by peak c. According to Hoffmann et al. [56], Auger-like process can occur in normal tunneling, causing the emission of ''forbidden'' photons whose energy exceed that of the tunneling electron, i.e. overbias emission [57,58].

Previous studies have demonstrated that the electric field $F_{FER}$ of FER states of orders $n$=1, 2 and 3 can be measured from the slope of the plot of FER energy $E_n$ versus $(n-1/4)^{2/3}$ [31,40], as shown in the inset in Fig. 1(a). At a constant current, $F_{FER}$ varies with tip sharpness. $F_{FER}$ is weaker as the tip is sharper [53]. Figure 3(a) shows FER spectra at different fields. When $F_{FER}$ increases, the separation $\Delta E_{10}$ between FER1 and FER0 peaks increases. Figure 3(b) depicts optical spectra with twin-peak features for bias voltages set at $E_1$ in Fig. 3(a). When $F_{FER}$ increases, the energy of peak b (marked by arrows) also increases. As shown in Fig. 3(c), the energies of peak b are close to the values of $\Delta E_{10}$, confirming that peak b results from the electron transition between the FER1 and FER0 states.

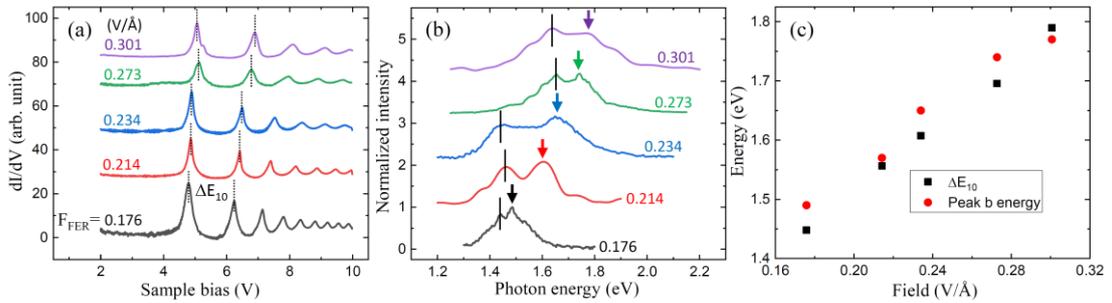

FIG. 3 (a) FER spectra of different electric fields ($F_{FER}$). $\Delta E_{10}$ is the separation energy between FER1 and FER0 peaks. (b) Optical spectra for bias voltages set at energies of FER1 peaks in (a). (c) $\Delta E_{10}$ values in (a) and energies of peaks marked by arrows in (b) versus $F_{FER}$.

As shown in Fig. 3(b), peak a (marked by lines) does not follow the tendency of peak b, resulting in splitting energy (SE) fluctuation in twin peaks. Because FER electrons have no angular momentum, the twin-peak feature is analogous to the hyperfine splitting of the 1s orbital of hydrogen atoms. This splitting is the result of the spin-spin interaction between the electron and the proton, whose SE is $\mu_0\hbar^2e^2g_eg_p/6\pi m_em_pa_0^3$ [59], where $g_e$ ($g_p$) and $m_e$ ($m_p$) are the g factor and mass of the electron (proton), respectively, $\mu_0$ is permeability constant, and $a_0$ is the Bohr radius. If we consider that SE depends on the mean distance $\bar{d}$ between the electron and the proton, which is $1.5a_0$ [60], the SE in our case equals to $9\mu_0\hbar^2e^2g_e^2/16\pi m_e^2\bar{d}^3$. Hence, the SE fluctuation observed in this experiment was due to the alterable mean distance between paired electron in the singlet state, and through this relation, $\bar{d}$ may fall between 0.22 and 0.34 Å for SE values in Fig. 3(b). However, SE values below 0.03 eV are difficult to observe because the individual linewidths of twin peaks are approximately 0.16 eV, which results in single peaks (see Supplemental Material [61]), indicating that $\bar{d}$ may exceed 0.4 Å. When the distance between two electrons reaches 0.22 Å, the Coulomb energy becomes 65.4 eV, which is considerably higher than the SE, implying that the exchange interaction can overcome Coulomb repulsion and force the two electrons to be close to each other at a balanced distance of 0.22 Å. Therefore, exchange force can be estimated from Coulomb force as 477 nN. SE is a quantity that can be used to examine the proximity of electrons in two-electron tunneling in the presence of exchange interaction and Coulomb repulsion.

As shown in Fig. 4(a), peak c in Fig. 1(c) remains almost constant when $F_{\text{FER}}$ changes, which differs from the scenario of peak b. The E state of peak c (hereinafter referred to as the Auger peak) is higher than the Fermi level of the STM tip. Therefore, we suggest that the potential $V_E$ for determining the E-state energy $E_e$ is a superposition of the external potential $V_x$ resulting from the bias voltage and image potential on the tip side $V_{\text{im}}$. Both $V_x$ and $V_{\text{im}}$ are dependent on tip sharpness, which can be quantified using an open angle θ in a conical model [Fig. 4(b)]. Our previous study demonstrated a method to secure θ from FER energies of high orders and their corresponding distances between the tip and the surface [53]. After acquiring θ, $V_x$ can be depicted. (see Supplemental Material [61]). The form of $V_{im}$ can be expressed as -e/β16πε₀s at the axis of symmetry of the cone, in which FER electrons move. In this form, s is the distance from the tip and β (>1) is a factor that describes a smaller $V_{im}$ compared with that in the plane case (β=1). In principle, β increases when θ decreases. According to the WKB approximation, $E_e$ can be calculated as follows

$$\int_0^d 0.512\sqrt{E_e - eV_E(z)}\, dz = n'\pi, \qquad (1)$$

where n' is an integer and d is the distance between the tip and the surface when the bias voltage is set to $E_0$. Thus, Auger peak is $E_e$–$E_0$. We used Eq. (1) to fit Auger peak to secure β. Figure 4(c) displays $V_e$ for Auger peak in Fig. 1(c), in which $V_x$ of θ=69° and $V_{im}$ of β=1.8 are depicted. Because of no solution for n'=1, n' for $E_e$ is 2. Figure 4(d) shows a plot of θ and β versus $F_{FER}$ for the Auger peaks in Fig. 4(a), indicating that while θ increases with increasing $F_{FER}$, β exhibits a reversal trend. Therefore, as expected, β is inversely proportional to θ [Fig. 4(e)].

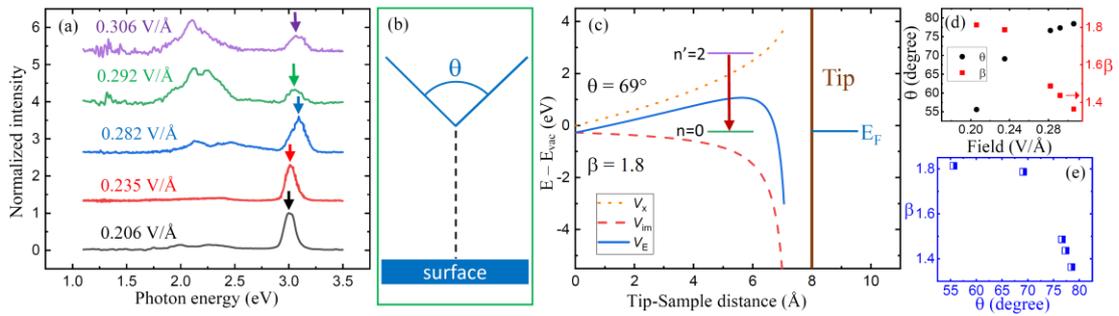

FIG. 4 (a) Optical spectra of bias voltages set at the energies of FER0 peaks under different electric fields. (b) The sharpness of the STM tip is defined by the open angle θ in the conical model. FER electrons move along the axis of symmetry of the cone, marked by a dashed line. (c) $V_E$ (solid curve), which is a superposition of $V_x$ (dotted curve) and $V_{im}$ (dashed curve), for the Auger peak in Fig. 1(c). In this case, θ=69° and β=1.8. (d) θ and β versus $F_{FER}$ for the Auger peaks in (a). (e) Plot of β versus θ.

Figure 5 shows a differential Z-V spectrum of both FER0 and FER1, revealing that the spectral intensities at the valleys (marked by triangles) around the FER1 peak are zero, which indicates an energy gap [40]. This zero valley intensity also indicates that the electrons in the FER1 state exhibit total reflection from the Ag(100) surface. Therefore, surface plasmons cannot be induced, and electron transition is the only channel available for electrons to exit the FER1 state. Although the FER0 state falls within the energy gap, Fig. 5 depicts a nonzero valley intensity on the left-hand side of the FER0 peak. This nonzero feature suggests that the electrons in the FER0 state may couple with surface electrons to generate surface plasmons to exit the FER0 state.

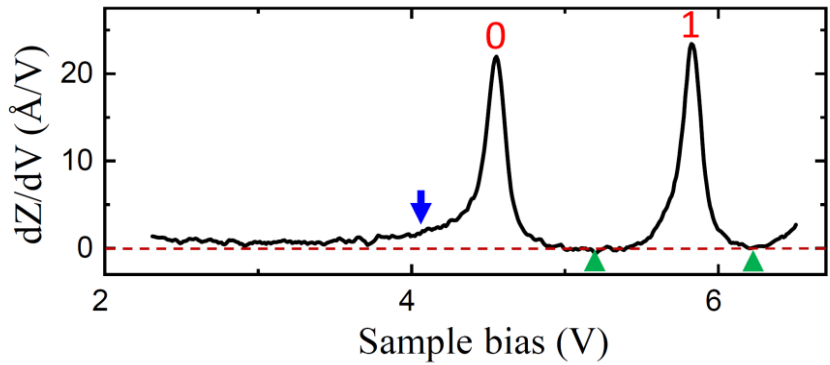

FIG. 5 Differential Z-V spectrum of FER0 and FER1. The dashed line indicates a zero spectral intensity. The spectral intensities at the valleys marked by triangles around the FER1 peak are zero. However, the valley intensity on the left-hand side of the FER0 peak, indicated by an arrow, is not zero.

In summary, light emission from FER states on Ag(100) can reveal dynamical processes of triplet-state formation and Auger-type excitation during the relaxation of paired FER electrons. Hence, optical spectra demonstrate twin-peak and Auger-peak characteristics. The Auger peak is obviously a single-photon source, which can be used for quantum information. Paired photons in twin-peak signals originate from triplet-state electrons. It is intriguing to investigate whether these two photons are connected by quantum entanglement. Technically, FER has potential to develop into a switchable light source by simply tuning the STM bias voltage.

The authors are grateful for the support provided by the Ministry of Science and Technology (Grant No: MOST 111-2112-M-001-086) and Academia Sinica (Grants No: AS-iMATE-109-15 and AS-iMATE-111-13), Taiwan.

**Supplemental Material:**

**Triplet State and Auger-Type Excitation Originating from Two-Electron Tunneling in Field Emission Resonance on Ag(100)**

Shin-Ming Lu, Ho-Hsiang Chang, Wei-Bin Su,[*] Wen-Yuan Chan, Kung-Hsuan Lin, and Chia-Seng Chang

*Institute of Physics, Academia Sinica, Nankang, Taipei 11529, Taiwan*

[*]Corresponding author.
wbsu@phys.sinica.edu.tw

## 1. Peak resulting from the relaxation of triplet-state electrons without splitting

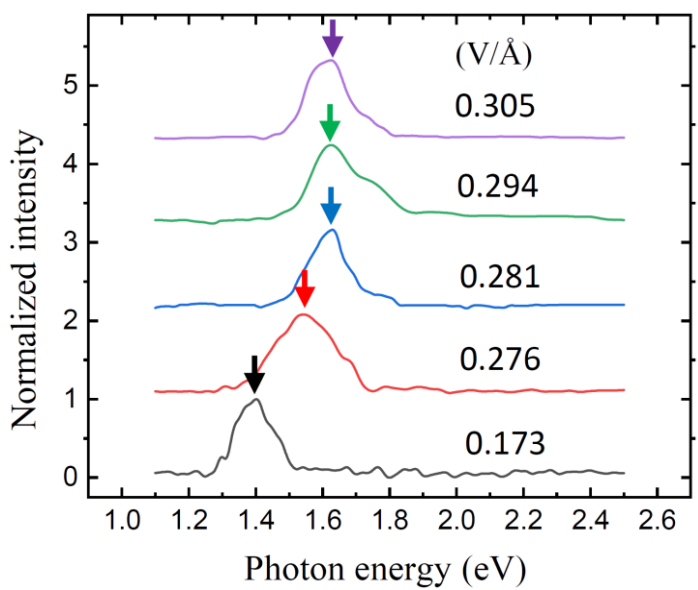

**Figure S1.** Optical spectra showing single peak under different electric fields. Peak moves toward a higher energy with increasing electric field, as indicated by arrows.

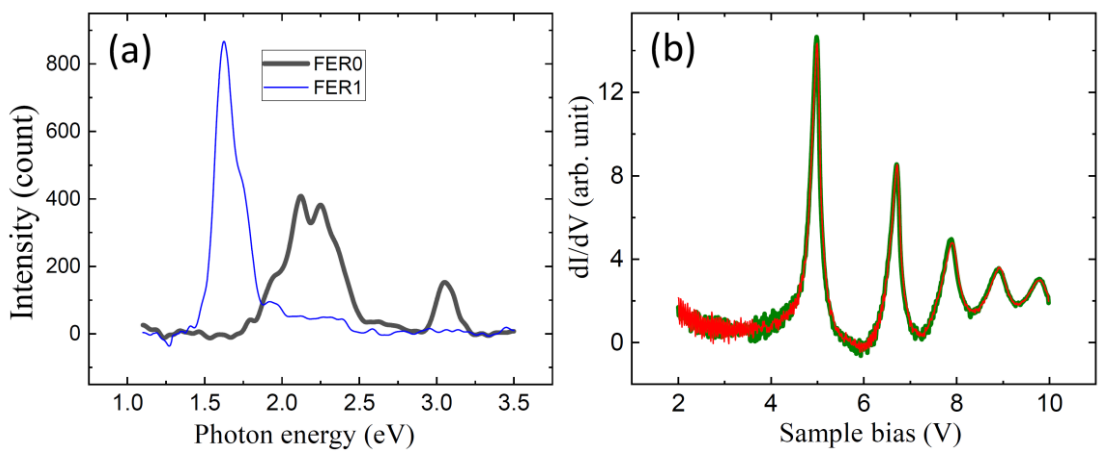

**Figure S2.** (a) Right after acquiring an optical spectrum (thin) showing the peak resulting from triplet-state electrons for the bias voltage set at $E_1$, the bias voltage was set at $E_0$ at once to take another optical spectrum (thick), which shows substantial changes including the appearance of the Auger peak and radiative decay of the surface plasmon. (b) Almost identical FER spectra before setting bias voltages at $E_1$ and $E_0$ to acquire optical spectra in (a), indicating that the tip was unchanged during the acquisition of two optical spectra in (a).

## 2. Details of obtaining $V_x$ for the bias voltage set at $E_0$

Based the method in Ref. [53], θ can be known, and it corresponds a constant ν. Figure S3 shows ν as a function of θ. Then $V_x(z)$ follows

$$V_x(z) = A(d-z)^\nu, \qquad (1)$$

where z is distance from the surface, and d is the distance between the tip and the surface when the bias voltage is set at $E_0$, and A is a coefficient. A can be obtained from

$$E_0 - W_{sample} + W_{tip} = Ad^\nu, \qquad (2)$$

where $W_{sample}$ and $W_{tip}$ are the work functions of the sample and tip, respectively. $W_{tip}$ is 5.5 eV for PtIr tip. $W_{sample}$ is obtained from the extrapolated value in the plot of $E_n$ versus $(n-1/4)^{2/3}$ like the inset in Fig. 1(a). $V_x(z)$ is depicted accordingly.

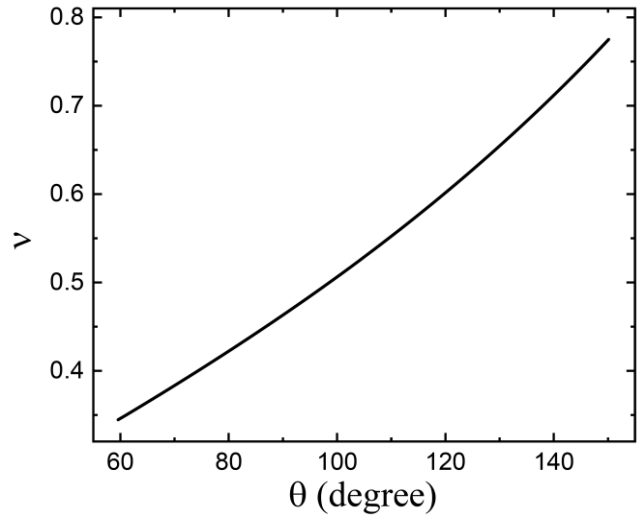

**Figure S3.** Plot of ν as a function of θ.